\iffalse
\documentclass{sig-alternate}
\else
\documentclass[10pt]{article}
\usepackage{amsthm, mdwlist} 
\fi

\usepackage{amsmath, amssymb, bbold, stmaryrd, mathpartir, cite, url}
\usepackage{mmm, reflective_grammars}

\title{Parsing Reflective Grammars\thanks{\small This research was made possible by the US National Science Foundation under grant number CCF-0811015, ``CPA-SEL: Developing a Theory of Hygienic Macros''.}}

\iffalse
\input{prg_ldta}
\else
\usepackage{authblk}
\author{Paul Stansifer}
\author{Mitchell Wand}
\affil{College of Computer and Information Science \\ Northeastern University\\ Boston, Massachusetts, United States
  \\ \{{\tt pauls},{\tt wand}\}{\tt @ccs.neu.edu}
  \footnote{A shorter version of this paper appeared in LDTA 2011.~\cite{ldtaversion}} }
\betterspacing
\fi

\begin{document}

\maketitle

\begin{abstract}
  Existing technology can parse arbitrary context-free grammars, but only a single, static grammar
  per input.  In order to support more powerful syntax-extension systems, we propose reflective
  grammars, which can modify their own syntax during parsing.  We demonstrate and prove the
  correctness of an algorithm for parsing reflective grammars.  The algorithm is based on Earley's
  algorithm, and we prove that it performs asymptotically no worse than Earley's algorithm on
  ordinary context-free grammars.
\end{abstract}

\section{Introduction}

A software project may involve many different languages with different purposes and complexities,
each with its own ``natural'' syntax.  Typically, these languages are segregated from each other,
either appearing in separate files, or inside strings.  But parenthesis-structured languages from
the Lisp family support incremental syntax extension (via macro systems).  This extension process
provides powerful integration, but the surface syntax is restricted to S-expressions.

We believe it is possible to bridge this gap and create macro systems with the syntactic power of
arbitrary context-free grammars.  However, new parsing technology is needed to do so.  In this
paper, we propose reflective grammars, which allow a language designer to define an incrementally
extensible base language.  In such a language, a valid sentence may contain strings matching
productions dynamically added by the sentence itself.  This happens in a structured fashion.  Users
of this language can use its extension construct to write in any surface syntax they want.

These language extensions are dynamic in the sense that they occur in the same file in which they
are used; they are structured in that they have well-defined scope; and they are recursive
in that an arbitrary number of extensions may be nested.

Our reflective grammars are based on context-free grammars.  Although many modern languages can be
made to fit into restricted subsets of context-free languages, such as LALR(1), context-free
languages are easier to understand and manipulate, and are closed under
composition~\cite{kats2010pure}.  This means that they are more suitable for languages which are to
be extended by the user.

Others have demonstrated impressive speed improvements to the Earley and GLR
algorithms~\cite{McLean1996,Aycock2001,Aycock2002a,Mandelbaum2009,McPeak2004}. We believe that the
historical performance motivations for using restricted subsets of context-free grammars no longer
apply.


A macro system could provide meaning to these syntactic extensions, but we do not present one here; this
paper only covers parsing.

In section \ref{sec:language}, we describe reflective languages in more detail.  Section
\ref{sec:algorithm} describes a recognition and parsing algorithm.  Section \ref{sec:complexity}
proves an upper bound to the time taken by parsing.  Sections \ref{sec:related} covers related work,
and section \ref{sec:conclusion} discusses our conclusion and future work.

\section{Reflective languages}
\label{sec:language}

\subsection*{Examples}
\label{sec:examples}

\newenvironment{arrowgrammar}{\begin{array}{l@{{}\rightarrow{}}l@{\quad}l}}{\end{array}}

The crux of our examples is the special right-hand side symbol $\mathbb{R}$.  In the grammar $G$,
the strings $w$ that $\mathbb{R}$ derives (denoted $\grs G {\mathbb{R}} w$), are the strings
in the set 

\[\{w_1w_2 : \grs G {\nt{Gram}} {w_1} \text{ and } \grs {G'} {S'} {w_2}\},\] 

where


\begin{itemize}
\item $\nt{Gram}$ is a distinguished nonterminal in $G$ such that strings derivable from \nt{Gram} can be interpreted as grammars by an operation denoted $\llbracket-\rrbracket$.
\item $G' = G \oplus \llbracket w_1 \rrbracket$, where $\oplus$ creates a new grammar by combining the productions of two grammars, and
\item $S'$ is the start symbol of $G'$.
\end{itemize}

For our examples, we will define a reflective grammar for a language containing numbers,
identifiers, and function invocations in the style of C-like languages.  In addition to these
conventional elements, the grammar accepts extensions, marked by pairs of curly
brackets.  The meaning of the extension symbol $\mathbb{R}$ depends on the nonterminal \nt{Gram},
which we also must define, giving a BNF-like meta-syntax for reflective grammars.  $\mathbb{R}$ is
represented in this notation as \term{REFL}.  The start nonterminal of the resulting grammar is
specified immediately after \term{gram}.

We assume
that the nonterminals \nt{Identifier}, \nt{Nonterm}, \nt{QuotedString}, and \nt{NaturalNumber}
have been given appropriate definitions already.  We also assume that whitespace is ignored, except
that \nt{Nonterm} and \nt{Identifier} follow standard tokenization rules.  Our parser
implementation successfully processes all the examples we give.

\[
\begin{arrowgrammar}
  \nt{Expr} & \nt{SimpleExpr} \term{(} \nt{Expr} \nt{MoreArgs} \term{)}
\\\nt{Expr} & \nt{SimpleExpr}
\\\nt{SimpleExpr} & \nt{Identifier}
\\\nt{SimpleExpr} & \nt{NaturalNumber}
\\\nt{SimpleExpr} & \verb|{{ | \mathbb{R} \verb| }}|
\\\nt{MoreArgs} &
\\\nt{MoreArgs} & \term{,} \ \nt{Expr} \nt{MoreArgs}
\\\nt{Gram} & \term{gram <} \nt{Nonterm} \term{> } \nt{Prods} \term{ end\_gram}
\\\nt{Prods} &
\\\nt{Prods} & \nt{Prod} \term{ } \nt{Prods}
\\\nt{Prod} & \term{<} \nt{Nonterm} \term{> ::= } \nt{RhsItems} \term{ ;}
\\\nt{RhsItems} &
\\\nt{RhsItems} & \term{<} \nt{Nonterm} \term{> }  \nt{RhsItems}
\\\nt{RhsItems} &  \nt{QuotedString}  \nt{RhsItems}
\\\nt{RhsItems} & \term{REFL }  \nt{RhsItems}
\end{arrowgrammar}\]

A simple sentence in the language of this grammar is \term{plus(1, plus(2,3))}.  A sentence that uses its
reflective capabilities to add simple infix operations is
\begin{verbatim}
plus(1, plus(2, 
          {{ gram <Expr>
               <Expr> ::= <SimpleExpr> <Op> <Expr> ;
               <Op> ::= "+" ;
             end_gram
             3 + plus(4, 5 + 6) }} ), 7)
\end{verbatim}

\sloppy

The extension recognizes the text between \term{gram} and \term{end\_gram} inclusive as being
derived from \nt{Gram}.  It interprets the grammar extension, and after that, it expects a string
derived from \nt{Expr} in the extended grammar, which it finds: \term{3 + plus(4, 5 + 6)}.  The
surrounding text, that is, \verb|plus(1, plus(2, {{| and \verb|}}), 7)|, is in the original grammar.
This means that the sentence

\fussy

\begin{verbatim}
plus(1, plus(2, 
          {{ gram <Expr>
               <Expr> ::= <SimpleExpr> <Op> <Expr> ;
               <Op> ::= "+" ;
             end_gram
             3 + plus(4, 5 + 6) }} ), 7 + 8)
\end{verbatim}
is \emph{not} in the grammar, because \term{7 + 8} is outside the $\mathbb{R}$ that provided a new
definition for \nt{Expr}.

Extensions can be used to gradually build up more powerful languages.  In the following example,
still in the same base grammar, we add lambda expressions and then infix operations (we represent 
$\lambda$ as \verb|\|, making the assumption that backslash is not already used as the escape 
character in string literals):
\begin{verbatim}
plus(1, 
  {{ gram <Expr>
       <Expr> ::= "\" <Identifier> "." <Expr> ;
       <SimpleExpr> ::= "(" <Expr> ")" ;
     end_gram
     (\x. plus(2,x))(
       plus(3,
         {{ gram <Expr>
              <Expr> ::= <SimpleExpr> <Op> <Expr> ;
              <Op>   ::= "+" ;
            end_gram
            (\y. 4 + y)(
              5 + (\z. 6 + z)(7)) }} )) }} )
\end{verbatim}

Note that the extension markers that this base grammar uses, \verb|{{ }}|, have no special status in
our system, and the user could choose to use them as another kind of delimiter, provided
he or she did so unambiguously.  The only reason they appeared in the base grammar at all because
omitting them would have made extensions hard to read, and even made it ambiguous where a grammar
extension ends after binary operations are permitted.

However, suppose that the author of the base language lacked this foresight, and had written the
extension rule as $\nt{SimpleExpr} \rightarrow \mathbb{R} $, instead of $\nt{SimpleExpr} \rightarrow
\verb|{{ | \mathbb{R} \verb| }}|$.  All would not be lost, because the user could have simply added
and then used a new, better construct using \term{REFL}, which represents the $\mathbb{R}$ construct
in our meta-syntax:

\begin{verbatim}
plus(1, gram <Expr>
          <Expr> ::= "{{" REFL "}}" ;
        end_gram
        {{ gram <Expr>
             <Expr> ::= <SimpleExpr> <Op> <Expr> ;
             <Op> ::= "+" ;
           end_gram
           2 + 3 }} )
\end{verbatim}

The old and now ambiguous extension syntax still remains, however.  This is because, for simplicity,
we have omitted from these examples the ability to remove productions from grammars.  It would be
very easy to add this, however.  Our formalism does not depend on any relationship between the
grammar being extended and the extension, but to obtain the complexity bounds of section
\ref{sec:complexity}, it must be possible to compute the extension quickly.

\subsection*{Definitions}

To define reflective grammars, we first need some metavariables.  Let $t$ range over terminal
symbols, $A$ and $B$ be nonterminals, $\a$, $\b$, $\g$, and $\delta$ be right-hand sides (strings of
terminals, nonterminals, and of the distinguished symbol $\mathbb{R}$), $x$ be the input string of
terminals, and let $i, j, k$, and $l$ be indices into that string.  We will use $x_{i,j}$ to
represent substrings of $x$.  The indices are zero-based and half-open; i.e., $x = x_{0,
  \left|x\right|}$.  The empty string will be represented with the symbol $\epsilon$.  We will name
other strings $w$.  Finally, we will use $G$ for a reflective grammar.

A reflective grammar $G$ consists of some set of productions $(A \rightarrow \a) \in G$, and a start
symbol $A = G.\text{start}$.

\subsection*{Semantics}
\label{sec:semantics}
In order to define the meaning of a reflective grammar, we must define the meaning of right-hand
sides.  We write $\grs G \a x$ to mean that the right-hand side $\a$ derives the string $x$
according to the grammar $G$. Right-hand sides are built recursively from terminals, nonterminals,
and the $\mathbb{R}$ symbol:

\begin{mathpar}
  \infer[L-Empty]{ }{\grs G \epsilon \epsilon} 
\\\infer[L-Terminal]{\grs G \a w}
{\grs G {\a t} {w t}}
\\\infer[L-Nonterminal]{\grs G \a {w_1} \\
                                (A\rightarrow\delta) \in G \\
                                \grs G \delta {w_2}}
                               {\grs G {\a A} {w_1 w_2}}
\\\infer[L-Reflection]
            {\grs G \a {w_1} \\
             \grs G {\nt{Gram}} {w_2} \\
             G' = G \oplus \llbracket w_2 \rrbracket\\ 
             (G'.\text{start} \rightarrow \delta) \in G' \\
             \grs {G'} \delta {w_3}}
            {\grs G {\a\mathbb{R}} {w_1 w_2 w_3}}
\end{mathpar}

We say $x \in L(G)$ (that is, $x$ is in the language of $G$), iff $\grs G {G.\text{start}} x$.

We restrict $\oplus$ by forbidding the user from extending the special \nt{Gram} nonterminal, and
the nonterminals that make it up, because the interpretation function $\llbracket-\rrbracket$ is
fixed, so it would not be able to interpret the newly-valid strings that \nt{Gram} derives.
However, a macro system using this parser could reasonably permit extensions to \nt{Gram} if the
user supplied a translation from the extended notation for grammars into the original notation.
Also, to make our complexity analysis simpler, we require that \nt{Gram} be non-nullable and appear
on the left-hand side of only one production.

\section{Recognizer algorithm}
\label{sec:algorithm}
We next present an algorithm for recognizing the language of a reflective grammar $G$, based on the
Earley recognizer algorithm~\cite{Earley1983}:

\begin{mathpar}
 \infer[R-Start]{G.\text{start} \rightarrow \delta \: \in G}
         {\ei 0 {G.\text{start}} {} {\delta} G \in S_0}
\\\infer[R-Shift]{\ei i A \a {t\b} G \in S_j \\ x_j = t}
                       {\ei i A {\a t} \b G \in S_{j+1}}
\\\infer[R-Call]{\ei i A \a {B\b} G  \in S_j \\  (B \rightarrow \delta) \in G}
                      {\ei j B {} \delta G  \in S_j}
\\\infer[R-Return]{\ei i A{\a}{B\b} G  \in S_j \\
                         \ei j B \delta {} G  \in S_k }
                        {\ei i A {\a B} \b G  \in S_k}
\end{mathpar}

\begin{mathpar}
\\\infer[R-Parse-grammar]{\ei i A \a {\mathbb{R}\b} G  \in S_j \\
                                  (\nt{Gram} \rightarrow \g) \in G}
                                 {\ei j {\nt{Gram}} {} \g G  \in S_j}
\\\infer[R-Refl-call]{\ei i A {\a}{\mathbb{R}\b} G  \in S_j  \\
                            \ei j {\nt{Gram}} {\g} {} G  \in S_k \\
                            G' = G \oplus \llbracket x_{j,k} \rrbracket \\
                            (G'.\text{start} \rightarrow \delta) \in G'}
                         {\ei{k}{G'.\text{start}}{}{\delta}{G'} \in S_k}
\\\infer[R-Refl-return]{\ei i A {\a} {\mathbb{R} \b} G  \in S_j \\
                              G' = G \oplus \llbracket x_{j,k} \rrbracket \\
                              \ei k {G'.\text{start}} {\delta} {} {G'} \in S_l}
                             {\ei i A {\a \mathbb{R}} {\b} G  \in S_l}
\end{mathpar}

An Earley recognizer accumulates Earley items.  An Earley item is a tuple $\ei i A \a \b G$, where
$(A \rightarrow \a\b) \in G$, and the cursor (the $\cdot$ symbol) marks a position in the right-hand
side $\a\b$.  The grammar $G$ is not part of traditional Earley items; we have added it for our
grammars. The algorithm collects sets $S_j$, where the set $S_j$ corresponds to the $j$th character
in the input string $x$.  The algorithm places the Earley item $\ei i A \a \b G$ in the set $S_j$
only if $\grs G \a x_{i,j}$.  However, for efficiency's sake, the recognizer only generates that
Earley item in the first place if it might be needed (the \textsc{R-Call} rule determines that a nonterminal might need
to be recognized at a particular point).

The recognizer proceeds strictly left-to-right.  The rules \textsc{R-Start} and \textsc{R-Call}
place items of the form ${\ei j A {} \delta G}$ in locations where the nonterminal $A$ is
expected to ``seed'' recognition of an $A$.  The \textsc{R-Shift} rule advances the cursor over an expected terminal.  The
\textsc{R-Return} rule advances the cursor over an expected nonterminal, provided there exists a
corresponding ``finished'' item of the form ${\ei j A \delta {} G}$.

The last three rules, \textsc{R-Parse-grammar}, \textsc{R-Refl-call}, and \textsc{R-Refl-return},
are our additions to the algorithm.  \textsc{R-Parse-grammar} and \textsc{R-Refl-call} are both
``seed'' rules, analogous to \textsc{R-call}.  \textsc{R-Parse-grammar} fires when the recognizer
reaches an $\mathbb{R}$, and it starts to consume a string matching \nt{Gram}.  When the \nt{Gram}
has been completely parsed, \textsc{R-Refl-call} creates an extended grammar, and descends into its start terminal.
Finally, \textsc{R-Refl-return} is analogous to \textsc{R-Return}; it is triggered by an Earley item
that indicates that a string matching the extended grammar is completed, and it advances the cursor
over the $\mathbb{R}$ that was waiting on it.

If $G' = G \oplus \llbracket x_{j,k} \rrbracket$, then we will say that $G'.\text{location} = (j,k)$
and $G'.\text{parent} = G$ (note that $G$ could be an extended grammar or just the base grammar).
We will compare grammars in an intensional fashion.  Two extended grammars will be equal exactly
when their locations and parents are the same, which implies that, in fact, they posses exactly the
same rules.  This will decrease the complexity of executing the \textsc{R-Refl-return} rule, and
make equality comparisons between Earley items fast.

The algorithm is considered to have recognized the string $x$ in the language $G$ iff it produces an
Earley item of the form ${\ei 0 {G.\text{start}} \delta {} G}$ in the last set, $S_{|x|}$.

\subsection*{Parsing instead of recognizing}
\label{sec:parsing}
There are two approaches to turn the recognizer into a parser.  If ambiguous parses are to be
rejected by the parser, Earley's simple technique suffices: In each Earley item, we associate each
nonterminal to the left of the cursor with a pointer to the ``completed'' Earley item $\ei j B
\delta {} G$ that derives it.  Items that have multiple pointers render any parse that uses them ambiguous.  

If a representation of all parses is desired, Scott's Buildtree algorithm~\cite{Scott} can be
adapted easily to our recognizer.  It depends on the recognizer annotating nodes with
``predecessor'' and ``reduction'' pointers.  Therefore, when a rule produces an item $\ei i A {\a\b} \gamma G \in S_k$,
where $\b$ is a single terminal, nonterminal, or $\mathbb{R}$, it adds a predecessor pointer from it to
the antecedent item $\ei i A \a {\b\gamma} G \in S_j$.  When \textsc{R-Return} produces $\ei i A {\a
  B} \gamma G \in S_k$, it adds a reduction pointer from it to the antecendent item $\ei j B \delta
{} G \in S_k$, and when \textsc{R-Refl-Return} produces a rule of the form $\ei i A {\a\mathbb{R}}
\gamma G \in S_k$, it adds a reduction pointer from it to the antecedent item $\ei j B \delta {}
{G'} \in S_k$.

Scott's algorithm traverses the Earley items and builds up a shared packed parse forest.  The symbol
nodes~\cite[p.~59]{Scott} are marked with a nonterminal and a beginning and ending position.  In a
reflective setting, these nodes must also have the grammar from which the nonterminal came, because
a nonterminal is only meaningful in the context of some grammar.

\subsection*{Correctness}
\iffalse
\newtheorem{correct}{Theorem}
\begin{correct}[Recognizer correctness]
\emph{ 
  \begin{center}
    $\grs G {G.\text{start}} x$ \\ 
    iff there is a production \\
    $(G.\text{start} \rightarrow \gamma) \in G \text{ such that } \ei 0 {G.\text{start}} {\gamma}
    {} {G} \in S_{|x|} $
  \end{center}
}
\end{correct}

We prove each direction separately.  The algorithm is complete:

\begin{center}
  $\grs G \a {x_{i,j}}$ and $\ei i A {} {\a\b} G \in S_i$ \\
  implies \\
  $\ei i A \a \b G \in S_j$
\end{center}

by induction on the derivation of $\grs G \a {x_{i,j}}$, and the algorithm is sound:
\[ \ei i A \a \b G \in S_j \text{ implies } \grs G \a x_{i,j} \]
by induction on the derivation of ${ \ei i A \a \b G \in S_j} $.

\hfill \\ 

The full proof is included in a longer version of this paper~\cite{longversion}.

\else







\newenvironment{pj}  {\begin{tabular}{p{5cm} p{5.5cm}}}  {\end{tabular}}

\newtheorem*{all-valid}{Completeness Lemma}
\newtheorem*{grammar-invariant}{Grammar Origin Lemma}
\newtheorem*{no-invalid}{Soundness Lemma}
\newtheorem*{correct}{Correctness}

Before we prove correctness, we present a slight reformulation of our semantics, where concatenation is
represented indirectly, by taking substrings of the input:

\begin{mathpar}
  \infer*[left=L-Empty]{ }{\grs G \epsilon \epsilon}
\\\infer*[left=L-Terminal]{\grs G \a {x_{i,j}} \\ x_{j} = t}
                          {\grs G {\a t} {x_{x,j+1}}}
\\\infer*[left=L-Nonterminal]{\grs G \a {x_{i,j}} \\
                                 (A\rightarrow\delta) \in G \\
                                \grs G \delta {x_{j,k}}}
                               {\grs G {\a A} {x_{i,k}}}
 \\\infer*[left=L-Reflection]
            {\grs G \a {x_{i,j}} \\
             \grs G {\nt{Gram}} {x_{j,k}} \\
             G' = G \oplus \llbracket x_{j,k} \rrbracket\\
              (G'.\text{start} \rightarrow \delta) \in G' \\
             \grs {G'} \delta {x_{k,l}}}
            {\grs G {\a\mathbb{R}} {x_{i,l}}}
\end{mathpar}

Proving the algorithm correct consists of two parts: that the algorithm recognizes all strings in
the language of the grammar (``valid strings'') and that it recognizes none that are not (``invalid
strings'').

\begin{all-valid}[the algorithm recognizes all valid strings]
\emph{ 
\[ {\grs G {G.\text{start}} x} \text{ implies } {\ei 0 {G.\text{start}} \delta {} G} \in S_{|x|}\]
}
\end{all-valid}
\begin{proof}
We will first show that, given an input string $x$ in the language of $G$,
\[ \grs G \a {x_{i,j}} \text{ and } \ei i A {} {\a\b} G \in S_i 
   \text{ implies } \ei i A {\a} {\b} G \in S_j\]

We will proceed by induction on the structure of the proof tree that $\grs G \a
{x_{i,j}}$.  Each case corresponds to a rule for generating right-hand sides
that recognize a string.

\begin{description}
\item \textsc{L-Empty}: $\a = \epsilon$.   $\grs G \epsilon x_{i,j}$ implies that $x_{i,j} = \epsilon$. This means that $i=j$.\\[0.6ex]
  Therefore, the item $\ei i A {} {\a\b} G$ is the same as the item $\ei i A {\a} {\b} G$, and it is
  already in $S_j$.
\item \textsc{L-Terminal}:  $\grs G \a {x_{i,j}}$ and $x_{j} = t$.\\[0.6ex]
  \begin{pj}
    $\ei i A {\a} {t \b} G \in S_{j}$, & by the induction hypothesis at $\grs G \a {x_{i,j}}$ \\[0.6ex]
    $\ei i A {\a t} {\b} G \in S_{j+1}$, & by \textsc{R-Shift}
  \end{pj}
  \item \textsc{L-Nonterminal}: $\grs G \a {x_{i,j}}$ and, for some
    $\delta$, $B \rightarrow \delta \: \in G$ and $\grs G \delta {x_{j,k}}$. \\[0.6ex]
    \begin{pj}
      $\ei i A {\a} {B \b} G \in S_j$, & by the induction hypothesis at $\grs G \a {x_{i,j}}$.\\[0.6ex]
      $\ei j B {} {\delta} G \in S_j$, & by \textsc{R-Call}. \\[0.6ex]
      $\ei j B {\delta} {} G \in S_k$, & by the induction hypothesis at $\grs G \delta {x_{j,k}}$. \\[0.6ex]
      $\ei i A {\a B} {\b} G \in S_k$, & by \textsc{R-Return}.
    \end{pj}
  \item \textsc{L-Reflection}: $\grs G \a {x_{i,j}}$ and $\grs G {\nt{Gram}} {x_{j,k}}$ and $G' = G \oplus \llbracket x_{j,k} \rrbracket$ and $\grs {G'} {G'.\text{start}} {x_{k,l}}$. \\[0.6ex]
    \begin{pj}
      $\ei i A {\a} {\mathbb{R}\b} G \in S_j$, & by the induction hypothesis at $\grs G \a {x_{i,j}}$. \\[0.6ex]
      $\ei j {\nt{Gram}} {} {\g} G \in S_j$, & by \textsc{R-Parse-grammar}, for all $(G \rightarrow \g) \in G$. \\[0.6ex]
      There is a $\g$ such that $\nt{Gram} \rightarrow \g \in G$ and  $\grs G \g {x_{j,k}}$, & by inversion of \textsc{L-Nonterminal}. \\[0.6ex]
      $\ei j {\nt{Gram}} \g {} G \in S_k$, & by the induction hypothesis at $\grs G \g {x_{j,k}}$ (which is higher up in the proof tree, so the induction hypothesis may be applied.) \\[0.6ex]
      
      Let $G'$ be $G \oplus \llbracket x_{i,j} \rrbracket$.\\[0.6ex]
      $\ei k {G'.\text{start}} {} \delta {G'} \in S_k$, & by \textsc{R-Refl-call}. \\[0.6ex]
      There is a $\delta$ such that $G'.\text{start} \rightarrow \delta \in G'$ and $\grs {G'} \delta {x_{k,l}}$, & by inversion of \textsc{L-Nonterminal}. \\[0.6ex]
      $\ei k {G'.\text{start}} \delta {} {G'} \in S_l$, & by the induction hypothesis at $\grs {G'} \delta {x_{k,l}}$. \\[0.6ex]
      $\ei i A {\a\mathbb{R}} {\b} G \in S_l$, & by \textsc{R-Refl-return}.
    \end{pj}
\end{description}

By the premise, $x_{0,|x|}$ (that is, $x$) is in the language of $G$, so, for some $\delta$, $\grs G
\delta x$.  By the \textsc{R-Start} rule, ${\ei 0 {G.\text{start}} {} \delta G} \in S_0$.  By the
above argument, we also know that ${\ei 0 {G.\text{start}} \delta {} G} \in S_{|x|}$, which is to
say that the algorithm has successfully recognized the string.

\end{proof}

\begin{grammar-invariant}[all extended grammars come from a parsed \nt{Gram}]
\ \\
\begin{center}
  \emph{For any extended grammar $G' = G \oplus \llbracket x_{j,k} \rrbracket$ that appears in an
    Earley item, there exists some Earley item $\ei j {\nt{Gram}} \g {} G \in S_k$.}
\end{center}
\end{grammar-invariant}

\begin{proof}
  By induction on the recognizer rules; only \textsc{R-Parse-grammar} creates new grammars, and it
  obeys the above condition.
\end{proof}




\begin{no-invalid}[the algorithm recognizes no invalid strings]
\emph{ 
\[  {\ei 0 {G.\text{start}} \delta {} G} \in S_{|x|} \text{ implies }  {\grs G {G.\text{start}} x} \]
}
 \end{no-invalid}
\begin{proof}

We will first show that, given a string $x$ that our algorithm recognizes as
being in the language of $G$,
\[ \ei i A {\a} {\b} G \in S_j \text{ implies } \grs G \a {x_{i,j}} \]

We will proceed by induction on the structure of the proof tree that $\ei i A
{\g} {\b} G \in S_j$.

\begin{description}
  \item \textsc{R-Shift}: $\ei i A {\a} {t\b} G \in S_{j-1}$ and $x_j = t$. \\[0.6ex]
    \begin{pj}
      $\grs G \a {x_{i,j-1}}$,  & by the induction hypothesis at $\ei i A {\a t} {\b} G \in S_{j-1}$. \\[0.6ex]
      $\grs G {\a t} {x_{i,j}}$, & by \textsc{L-Terminal}.
    \end{pj}
  \item \textsc{R-Return}: $\ei i A {\a} {B \b} G \in S_j$ and $\ei j B {\delta} {} G \in S_k$. \\[0.6ex]
    \begin{pj}
      $\grs G \a {x_{i,j}}$, &by the induction hypothesis at $\ei i A {\a} {B \b} G \in S_j$ \\[0.6ex]
      $\grs G \delta {x_{j,k}}$, & by the induction hypothesis at $\ei j B {\delta} {} G \in S_k$. \\[0.6ex]
      $(B \rightarrow \delta) \in G$, & by the definition of Earley items. \\[0.6ex]
      $\grs G {\a B} {x_{i,k}}$, & by \textsc{L-Nonterminal}.
    \end{pj}
  \item \textsc{R-Refl-return}: $\ei i A {\a} {\mathbb{R}\b} G \in S_j$ and $G'= G \oplus \llbracket x_{j,k} \rrbracket$ and $\ei k {G'.\text{start} } {\delta} {} {G'} \in S_l$. \\[0.6ex]
    \begin{pj}
      $\grs G \a {x_{i,j}}$, & by the induction hypothesis at $\ei i A {\a} {\mathbb{R}\b} G \in S_j$. \\[0.6ex]
      $\ei j {\nt{Gram}} \g {} G \in S_k$, &by the Grammar Origin Lemma. \\[0.6ex]
      $\grs G {\nt{Gram}} {x_{i,k}}$, & by the induction hypothesis at $\ei j {\nt{Gram}} {\g} {} G \in S_k$. \\[0.6ex]
      $\grs {G'} \delta {x_{k,l}}$, & by the induction hypothesis at $\ei k {G'.\text{start} } {\delta} {} {G'} \in S_l$. \\[0.6ex]
      $G'.\text{start} \rightarrow \delta \in G'$, & by the definition of Earley items.\\[0.6ex]
      $\grs G {\a \mathbb{R}} {x_{i,l}}$, & by \textsc{L-Reflection}.
    \end{pj}
  \item All remaining rules produce Earley items of the form $\ei i A {} {\delta} G \in S_i$.  
    \begin{pj}
      $\grs G \epsilon \epsilon$, & by \textsc{Empty}.
    \end{pj}

\end{description}

Therefore, since the algorithm produced an Earley item of the form ${\ei i {G.\text{start}} \delta
  {} G}$ in the set $S_{|x|}$, we know that $\grs G \delta {x_{0,|x|}}$.  Because $(G.\text{start}
\rightarrow \delta) \in G$, we know that $x$ is in the language of $G$.

\end{proof}

\begin{correct}[the algorithm is correct]
\emph{ 
\[ {\grs G {G.\text{start}} x} \quad \text{ iff } \quad  {\ei 0 {G.\text{start}} \delta {} G} \in S_{|x|}\]
}
\end{correct}

\begin{proof}
  By the soundness and completeness lemmas above, the algorithm recognizes a string iff it is valid.
\end{proof}


\fi

\section{Complexity}
\label{sec:complexity}
We will characterize the complexity of this algorithm in terms of both the length of the input
string and the nature of extended grammars it defines.  Let $n$ be the length of the input string,
and let $g$ be the maximum size of any extended grammar defined.  We define the size of a grammar to
be the sum of the number of productions and the length of the right-hand sides.  By this definition,
there are only $g$ distinct values of $\rulepos A \a \b$ possible in a grammar of size $g$.

At each input position, there is some set of grammars which might be the current grammar,
given the part of the string to the left of the character.  Let $m$ be the maximum of the size of
these sets, over the length of the string.  Having $m$ be greater than 1 occurs in cases where
something else shares syntax with a syntax extension construct, or when the extension is not
terminated unambiguously, both of which are undesirable in practice.  However, in pathological
cases, $m$ grows exponentially with $n$. We know $m$ is always finite because grammar extensions are
applied in the order encountered and \nt{Gram} is non-nullable, so every grammar is uniquely defined
by sequence of distinct nonoverlapping nonempty substrings of the input string.  It is possible to
limit the value of $m$ and abort parsing if it exceeds some preset value.

Before we proceed, we must specify the behavior of $\llbracket x \rrbracket$ and $\oplus$.  We require
that both of those take no more than $O(ngm)$ time.  Most natural definitions will satisfy this
easily, as the string $x$ is no more than $n$ characters long, and the grammars produced by $\oplus$ and $\llbracket x \rrbracket$ have size no more than $g$.

Now we shall prove that recognition takes $O(n^3g^3m^3)$ time.  Our argument follows that of Earley~\cite{Earley1983}.

First, we observe that the algorithm can be executed by first determining the contents of $S_0$,
then $S_1$, and so on, because the contents of each $S$ never depends on an $S$ further to the
right.  Furthermore, every rule that places an Earley item into set $S_i$ has as an antecedent the
existence of an Earley item in $S_i$, with the exception of \textsc{R-Start} and \textsc{R-Shift}.
Imagining for the moment that each $S_i$ is a set that allows mutation by adding members, we sketch
out a strategy for taking the closure of our rules:

For each $S_i$, in order, ``seed'' the set by executing \textsc{R-Start} if $i=0$, or
\textsc{R-Shift} on every appropriate item in $S_{i-1}$ otherwise.  Now close the set over the
remaining rules: Apply all rules to the new Earley items, the result of which becomes the new Earley
items for the next iteration, repeating until no new items appear.

This closure process is the heart of the algorithm.  For each Earley item generated, it will execute
the rules, and insert the resulting item (if any) into the appropriate set.  There is one set of
Earley items for each input character, so the asymptotic running time is

\begin{center}
  number-of-input-characters $\times$ number-of-Earley-items-per-set $\times$ (rule-execution-time
  $+$ items-produced-per-item $\times$ set-insertion-time).
\end{center}

There are $n$ input characters.  Each set contains at most $O(ngm)$ Earley items: in the form
$\ei i A \a \b {G_1}$, there are $n$ possible values of $i$, $g$ possible values for $\rulepos A \a
\b$, and
 the number of distinct grammars $G_1$ in the set is limited to $m$.

 If each set is represented as an array of length $n$ containing linked lists of items, and an item
 anchored at $i$ is stored in the list at index $i$ of the array, there will be at most $O(gm)$
 items in each linked list.  To perform set insertion by adding elements to these lists, we also
 need to compare Earley items for equality quickly.  It is possible to store all the components of
 our Earley items as indices for constant-time comparison.  This is trivial for the anchor $i$ and
 for the rule position $\rulepos A \a \b$, but requires explanation for the grammar $G$.  The
 contents of grammars can be stored in a table, and each Earley item's reference to the current
 grammar can be stored as an index into that table.  We have required that there only be one
 production of the form $\nt{Gram} \rightarrow \g$, so for each grammar with location $(i,j)$ and
 parent $G'$, there is only one possible Earley item that can produce it via \textsc{R-Refl-call}.
 This means that newly created grammars are unequal to all existing grammars, so the table never
 needs to be searched.  Therefore, comparing Earley items to each other takes constant time, and
 therefore inserting an Earley item into the set $S_i$ takes $O(gm)$ time.

 Now, all that remains is to determine, per input item, how long the rules take to execute, and how
 many items the rule produces. Each rule (other than \textsc{R-Start}, which takes $O(g)$ time to
 execute overall) has at least one Earley item as a antecedent.  To apply the rule to an Earley
 item, we substitute the item into the antecedent, and then test the remaining antecedents.  This
 means that rules with two Earley items as antecedents will be attempted twice and succeed the
 second time.

\begin{description}
\item \textsc{R-Shift}  This rule takes $O(1)$ time to test the expected terminal against the input
  string.  It produces at most a single item.
\item \textsc{R-Call}  This rule needs to walk $G$, so it takes $O(g)$ time, producing at most
  $O(g)$ items.
\item \textsc{R-Return}  We reproduce the rule below:

\[ \\\infer[R-Return]{\ei i A{\a}{B\b} G  \in S_j \\
                         \ei j B \delta {} G  \in S_k }
                        {\ei i A {\a B} \b G  \in S_k} 
\]

We will show that the rule takes $O(ngm)$ time and produces $O(ngm)$ items.  It is always true that
$j \le k$, because the end of a production must not come before its start.  There are two possible
ways that an Earley item could be relevant to this rule:\footnote{Here, we differ from Earley by
  omitting a small optimization; he only tests items for applicability as the $\ei j B \delta {} G$
  antecedent in the \textsc{R-Return} rule.  This always works when $j < k$, and sometimes works
  when $j=k$.  Additional work must be done to make this behave correctly in the presence of
  nullable productions.  Aycock~\cite{Aycock2002a} discusses three different solutions to this problem.}

  If we have the item $\ei j B \delta {} G \in S_k$\footnote{An anonymous reviewer points out that the value of $\delta$ is irrelevant in executing this rule.  therefore, an intermediate rule could collapse all items of the form $\ei i B \delta {} G \in S_k$ into a special item $(i, B \rightarrow \square, G) \in S_k$, which the \textsc{R-return} rule could look for instead, reducing the number of times it executes.  However, this would not have an asymptotic effect on performance; the number of distinct possible values of $B \rightarrow \square$, like the number of distinct possible values of $B \rightarrow \delta \cdot$, is in $O(g)$.}, we know what $j$ is and that all matching items
  are in $S_j$.  There are $O(ngm)$ items in $S_j$ which need to be checked to see if they match
  $\ei i A \a {B\b} G$.  All of them could match: this rule could produce as many as $O(ngm)$ items.

  But if we have the item $\ei i A{\a}{B\b} G \in S_j$, the only matching Earley items that
  could have already been produced are those for which $j=k$.  So, we need to search $S_j$, which
  takes $O(gm)$ time to produce $O(gm)$ items, because the anchor of the item we are looking for is
  known to be $j$. The fact that $S_j$ is only partially complete at this point is of no
  consequence; whichever item arrives last in $S_j$ will succeed in finding the other.
\item \textsc{R-Parse-grammar} Like \textsc{R-Call}, this takes $O(g)$ time, producing at most
  $O(g)$ items.
\item \textsc{R-Refl-call}  Computing $G \oplus \llbracket x_{j,k} \rrbracket$ takes $O(ngm)$ time,
  as specified above.  \nt{Gram} is required to be non-nullable, so $j < k$, and therefore %
  the $\ei j {\nt{Gram}} \g {} G \in S_k$ item always appears last.  Searching $S_j$ for items
  matching $\ei i A \a {\mathbb{R}\b} G$ takes $O(ngm)$ time and produces at most $O(ngm)$ items.
\item \textsc{R-Refl-return}  $G'.\text{location} = (j,k)$, and $G'.\text{parent} = G$.  Other than
  that extra bookkeeping, this rule proceeds like \textsc{R-Return}.
\end{description}

For each Earley item, executing the rules takes $O(ngm)$ time and produces up to $O(ngm)$ items.
Each item that is produced needs to be inserted into the appropriate set (which, as we saw above,
takes $O(gm)$ time).  The deduplication performed by set insertion ensures we only have to execute
the rules once per \emph{unique} Earley item, even if the item is produced multiple times.
Otherwise, execution time would be slower, and it would even diverge in the case of left-recursive
rules.

Our total running time therefore is $n \times O(ngm) \times ( O(ngm) + O(ngm) \times O(gm)) =
O(n^3g^3m^3)$.  If the rules \textsc{R-Parse-grammar}, \textsc{R-Refl-call}, and
\textsc{R-Refl-return} are omitted, the original Earley algorithm is recovered.  The
\textsc{R-Return} rule, which remains, can still take $O(ngm)$ time and produce $O(ngm)$ items, so
the complexity is the same without the reflective rules.  Since Earley supports a single grammar of
fixed size, $g$ and $m$ are constants.  This is consistent with Earley's $O(n^3)$ result.  Our
system is therefore ``pay-as-you-go'': its reflective features have no asymptotic cost if they are
not used.

Earley recognition provides further performance guarantees in cases where the input obeys certain
restrictions.  We have not examined whether those same guarantees apply to our work.

\subsection*{Buildtree complexity}
The Buildtree algorithm of Scott~\cite{Scott}, introduced in section \ref{sec:parsing}, can be used
to construct parse trees (based on Earley items) when the results of ambiguous parses are needed in
a compact format.  (An ambiguous grammar may parse a sentence exponentially many or even infinitely many
ways.)

Scott's complexity analysis asserts that Buildtree takes time proportional to

\begin{center}
  number-of-input-characters $\times$ number-of-Earley-items-per-set $\times$ predecessor-items-per-item
\end{center}

The number of predecessor items an Earley item may have, as in Scott's work, is $n$.  To see this,
observe that an item where the cursor follows a nonterminal, \[\ei i A {\a B} \b G \in S_j\] can have as predecessor any item of the form \[\ei i A \a {B \b} G \in S_k\] where $0 \le k \le j$.  This same argument applies to cases where the cursor follows a $\mathbb{R}$.

On the other hand, if the cursor follows a terminal, there is
exactly one predecessor, and items where the cursor is at the beginning of the right-hand side have
no predecessor.

The number of input characters is $n$.  As above, the number of Earley items in each of our sets is
$O(ngm)$.  So executing Buildtree requires $O(n^3gm)$.  This means that Buildtree, which takes place
only once (after recognizing is completed), requires less time than recognizing, so it does not
affect the overall complexity.

\section{Related work}
\label{sec:related}
\subsection*{Parsers}

The idea of modifying an Earley parser to parse a more powerful class of grammars was inspired by
YAKKER~\cite{Jim2010a}, a powerful Earley-based parser for dependent grammars.  A dependent grammar
can, for example, recongize the language of strings containing a literal number $n$ followed by
a sequence of precisely $n$ characters.

Derivative-based parsing~\cite{Might2010} is an approach to parsing context-free languages in which
the parse state at a given character is simply a grammar representing the language of strings that
are valid suffixes to the already-parsed portion. The authors suggest that it could be used to
implement reflective grammars, but supply no details.

Like context-free grammars, parsing expression grammars (PEGs) can be composed by combining
productions to produce a legal grammar~\cite{Ford04}.  However, the ordered choice provided by PEGs
is not a true union, and ``incorrect orderings can cause suble errors''~\cite{kats2010pure}.  For
example, adding an \term{if}\ldots\term{then} construct can turn an existing
\term{if}\ldots\term{then}\ldots\term{else} construct into a syntax error.

\subsection*{Language extension systems}
There are a variety of systems that tackle the issue of syntax extensibility.  Each work in this
category is a complete system that tackles both the issue of parsing and the issue of
transformation.  We will only cover the comparable portion here, the parsers.  

A few of these systems parse input using some kind of dynamic grammars which, like ours, support
multiple grammars in one file.

Kolbly~\cite{Kolbly2002} describes a syntax extension system with an Earley-based parser that can
parse different regions of a file in different grammars.  However, all grammar extensions must be
predefined by the language designer --- the user cannot extend the language.

Another macro system with flexible syntax is ZL~\cite{Atkinson2010}.  It allows new syntax to be
added to C, though a system of iterated re-parsing.  However, it restricts what syntactic forms the
user may add.

Although Dylan's macro system~\cite{Bachrach1999} does not involve any special parser technology, it
does loosen Lisp's parentheses to a ``syntactic skeleton'', giving macro authors more control over
the appearance of macro invocations.

Gel~\cite{falcon2009gel} is a language syntax that, by requiring adherence to whitespace
conventions, correctly parses code that looks like Java, CSS, Smalltalk, and ANTLR.  Their goal is
in some ways a mirror image of ours: they unify a set of existing syntaxes into one large syntax,
while we describe how a single small syntax can be extended into many others in the bounds of one
file.

The Silver project~\cite{VanWyk2008} is a system for describing and extending languages, and
transforming those languages using attribute grammars.  Schwerdferger and Van Wyk
describe~\cite{Schwerdfeger2009} a static analysis for language extensions which ensures that, given
a host language, any number of these extensions can be added to the host language, and the result
will be LALR(1), as their parser requires.  However, they must significantly restrict the
permissible forms of syntax extensions in order to do so.

Metafront~\cite{Brabrand2003} is a system for defining languages and transformations between them.
They describe a novel type of grammar called a ``specificity grammar''.  In such a grammar, more
specific productions have priority over less specific productions.  Although composing their
grammars can produce errors, these errors can be expressed entirely in terms of the productions
involved, rather than as confusing shift/reduce and reduce/reduce conflicts. They also have what
they describe as a macro system; however, their macro definitions always have the scope of an entire file,
so they can use existing parser technology.

A system described by Cardelli, Matthes, and Abadi~\cite{Cardelli1994} discusses incrementally
extending grammars by adding productions (and grammar restriction, where productions are removed).
It rejects compositions of grammars that are not LL(1), but provides powerful integration between
grammar definitions and transformations.

Camlp4~\cite{DeRauglaudre2003} is a preprocessor for the Ocaml language.  It allows the user to
extend the Ocaml syntax.  It allows the language designer to select what parser the resulting,
extended, language will be parsed with, but the user must select one language per file.

\section{Conclusion and future work}
\label{sec:conclusion}

We have defined a class of grammars that specify languages that can modify their own
syntax during parsing.  We have presented an algorithm that can parse these reflective grammars and can parse
nonreflective grammars as fast as an ordinary Earley parser.  Furthermore, we have placed bounds 
on how costly the reflective feature is, in terms of how it is used.

We intend this work as the first step in building a macro system applicable to languages that lack
parenthesis-based syntax.  Our next steps will be to define requirements for a powerful and usable
macro system, and describe how such a macro system would interact with this parser.  In such a
system, there would be no special syntax for macro invocation, so user-defined syntax would be
indistinguishable from core syntax.  With the dynamic power of our parser, it would be possible to
have local definitions for macros, and even to import macros in a restricted scope.


\bibliography{parsing_reflective_grammars}{}

\begin{thebibliography}{10}

\bibitem{Atkinson2010}
K.~Atkinson, M.~Flatt, and G.~Lindstrom.
\newblock {ABI} compatibility through a customizable language.
\newblock {\em Proceedings of the Ninth International Conference on Generative
  Programming and Component Engineering - GPCE '10}, page 147, 2010.

\bibitem{Aycock2002a}
J.~Aycock.
\newblock Practical {E}arley parsing.
\newblock {\em The Computer Journal}, 45(6):620--630, June 2002.

\bibitem{Aycock2001}
J.~Aycock and N.~Horspool.
\newblock Directly-executable {E}arley parsing.
\newblock In R.~Wilhelm, editor, {\em Compiler Construction}, volume 2027 of
  {\em Lecture Notes in Computer Science}, pages 229--243. Springer Berlin /
  Heidelberg, 2001.
\newblock 10.1007/3-540-45306-7\_16.

\bibitem{Bachrach1999}
J.~Bachrach and K.~Playford.
\newblock D-expressions: {L}isp power, {D}ylan style.
\newblock \url{http://people. csail.mit.} \url{edu/ jrb /Projects /dexprs.htm},
  1999.

\bibitem{Brabrand2003}
C.~Brabrand, M.~I. Schwartzbach, and M.~Vanggaard.
\newblock The metafront system: Extensible parsing and transformation.
\newblock {\em Electronic Notes in Theoretical Computer Science},
  82(3):592--611, Dec. 2003.

\bibitem{Cardelli1994}
L.~Cardelli, F.~Matthes, and M.~Abadi.
\newblock Extensible syntax with lexical scoping.
\newblock \url{http://lucacardelli. name/ Papers/SRC-121.ps}, 1994.

\bibitem{DeRauglaudre2003}
D.~de~Rauglaudre.
\newblock {C}amlp4 - reference manual.
\newblock \url{http://caml.inria.fr/pub/docs/manual-camlp4/}, 2003.

\bibitem{Earley1983}
J.~Earley.
\newblock An efficient context-free parsing algorithm.
\newblock {\em Communications of the ACM}, 26(1), 1970.

\bibitem{falcon2009gel}
J.~Falcon and W.~Cook.
\newblock Gel: A generic extensible language.
\newblock In {\em Domain-Specific Languages}, pages 58--77. Springer, 2009.

\bibitem{Ford04}
B.~Ford.
\newblock Parsing expression grammars: a recognition-based syntactic
  foundation.
\newblock In {\em Proceedings ACM Symposium on Principles of Programming
  Languages}, pages 111--122, 2004.

\bibitem{Jim2010a}
T.~Jim, Y.~Mandelbaum, and D.~Walker.
\newblock Semantics and algorithms for data-dependent grammars.
\newblock {\em Annual Symposium on Principles of Programming Languages}, 45(1),
  2010.

\bibitem{kats2010pure}
L.~Kats, E.~Visser, and G.~Wachsmuth.
\newblock Pure and declarative syntax definition: Paradise lost and regained.
\newblock {\em Proceedings of Onward! 2010}, 2010.

\bibitem{Kolbly2002}
D.~M. Kolbly.
\newblock {\em Extensible Language Implementation}.
\newblock Ph.{D.}, University of Texas at Austin, 2002.

\bibitem{Mandelbaum2009}
Y.~Mandelbaum and T.~Jim.
\newblock Efficient {E}arley parsing with regular right-hand sides.
\newblock {\em Workshop on Language Descriptions Tools and Applications}, 2009.

\bibitem{McLean1996}
P.~McLean and R.~Horspool.
\newblock A faster {E}arley parser.
\newblock In T.~Gyim\'othy, editor, {\em Compiler Construction}, volume 1060 of
  {\em Lecture Notes in Computer Science}, pages 281--293. Springer Berlin /
  Heidelberg, 1996.
\newblock 10.1007/3-540-61053-7\_68.

\bibitem{McPeak2004}
S.~McPeak and G.~C. Necula.
\newblock Elkhound: A fast, practical {GLR} parser generator.
\newblock {\em Compiler Construction}, 2004.

\bibitem{Might2010}
M.~Might and D.~Darais.
\newblock Yacc is dead.
\newblock \url{http://arxiv.org/abs/1010.5023}, Oct. 2010.

\bibitem{Schwerdfeger2009}
A.~C. Schwerdfeger and E.~R. {Van Wyk}.
\newblock Verifiable composition of deterministic grammars.
\newblock {\em Conference on Programming Language Design and Implementation},
  44(6), 2009.

\bibitem{Scott}
E.~Scott.
\newblock {SPPF}-style parsing from {E}arley recognisers.
\newblock {\em Electron. Notes Theor. Comput. Sci.}, 203:53--67, April 2008.

\bibitem{ldtaversion}
P.~Stansifer and M.~Wand.
\newblock Parsing reflective grammars.
\newblock {\em LDTA}, 2011.
\newblock To appear.

\bibitem{VanWyk2008}
E.~R. {Van Wyk}, D.~Bodin, J.~Gao, and L.~Krishnan.
\newblock {S}ilver: an extensible attribute grammar system.
\newblock {\em Electronic Notes in Theoretical Computer Science},
  203(2):103--116, Apr. 2008.

\end{thebibliography}
\bibliographystyle{abbrv}

\end{document}